\documentclass[12pt,preprint]{aastex}

\newcommand\HII{H\,{\sc ii}}

\newcommand\HI{H\,{\sc i}}

\newcommand\kms{km~s$^{-1}$}

\newcommand\Msun{M$_{\odot}~$}

\newcommand\cmthree{cm$^{-3}~$}
\newcommand\cmsix{cm$^{-6}~$}

\newcommand\etal{et al.~}
\newcommand\be{\begin{equation}}
\newcommand\ee{\end{equation}}
\newcommand\bea{\begin{eqnarray}}
\newcommand\eea{\end{eqnarray}}

\newcommand\ddeg{$^{o}$}

\catcode`\@=11
\newcommand{\gsim}{${\mathrel{\mathpalette\@versim>}}$}
\newcommand{\lsim}{${\mathrel{\mathpalette\@versim<}}$}
\newcommand{\@versim}[2]{\lower 2.9truept \vbox{\baselineskip 0pt \lineskip
    0.5truept \ialign{$\m@th#1\hfil##\hfil$\crcr#2\crcr\sim\crcr}}}
\catcode`\@=12

\slugcomment{Accepted by ApJ}

\shorttitle{NGC 2024 (Orion B, W12): PDR Properties \& Magnetic field } 
\shortauthors{Roshi, Goss \& Jeyakumar}

\begin{document}

\title{VLA and GBT Observations of Orion B (NGC 2024, W12) : 
Photo-dissociation Region Properties and Magnetic field}

\author{D. Anish Roshi}
\affil{National Radio Astronomy Observatory\altaffilmark{1}, Charlottesville \& Green Bank,
520 Edgemont Road, Charlottesville, VA 22903, USA; aroshi@nrao.edu}

\author{W. M. Goss}
\affil{National Radio Astronomy Observatory, P.O. Box O, Socorro, NM 87801, USA;
mgoss@nrao.edu}

\author{S. Jeyakumar}
\affil{
Departamento de Astronom{\'i}a, Universidad de Guanajuato, AP 144, Guanajuato CP 36000, Mexico;
sjk@astro.ugto.mx
}

\altaffiltext{1}{The National Radio Astronomy Observatory is a facility of
the National Science Foundation operated under a cooperative
agreement by Associated Universities, Inc.}

\begin{abstract}
We present images of C110$\alpha$ and H110$\alpha$ radio 
recombination line (RRL) emission at 4.8 GHz and images of 
H166$\alpha$, C166$\alpha$ and X166$\alpha$ RRL emission at 1.4 GHz,
observed toward the starforming region NGC 2024.
The 1.4 GHz image with angular resolution $\sim$ 70\arcsec\ is 
obtained using VLA data. The 4.8 GHz image with angular resolution
$\sim$ 17\arcsec\ is obtained by combining VLA and GBT data
in order to add the short and zero spacing data in the uv plane. These images 
reveal that the spatial distributions of C110$\alpha$ line emission 
is confined to the southern rim of
the \HII\ region close to the ionization front whereas the C166$\alpha$ line
emission is extended in the north-south direction across the \HII\ region. 
The LSR velocity of the C110$\alpha$ line is 10.3 \kms\ similar to that of 
lines observed from
molecular material located at the far side of the \HII\ region. This similarity 
suggests that the
photo dissociation region (PDR) responsible for C110$\alpha$ line emission is 
at the far side of the \HII\ region. The LSR velocity of C166$\alpha$
is 8.8 \kms. This velocity is comparable with the velocity of
molecular absorption lines observed from the foreground gas, suggesting that
the PDR is at the near side of the \HII\ region. Non-LTE models
for carbon line forming regions are presented. Typical properties of the
foreground PDR are $T_{PDR} \sim 100$ K, $n_e^{PDR} \sim 5$ \cmthree,
$n_H \sim 1.7 \times 10^4$ \cmthree, path length $l \sim 0.06$ pc 
and those of the far side PDR are $T_{PDR} \sim$ 200 K, $n_e^{PDR} \sim$ 50 \cmthree,
$n_H \sim 1.7 \times 10^5$ \cmthree, $l \sim$ 0.03 pc. 
Our modeling indicates that the far side PDR is located within the \HII\ region.
We estimate magnetic field strength in
the foreground PDR to be 60 $\mu$G and that in the far side PDR 
to be 220 $\mu$G. Our field estimates compare well with the values 
obtained from OH Zeeman observations toward NGC 2024.
The H166$\alpha$ spectrum shows narrow (1.7 \kms)
and broad (33 \kms) line features. 
The narrow line has spatial distribution and central velocity ($\sim$ 9 \kms) similar
to that of the foreground carbon line emission, suggesting that they are associated.
Modeling the narrow H166$\alpha$ emission provides physical properties $T_{PDR} \sim 50 $ K,
$n_e^{PDR} \sim 4$ \cmthree and $l \sim 0.01$ pc and implies an ionization fraction of 
$\sim$ 10$^{-4}$. 
The broad H166$\alpha$ line originates from the \HII\ region.
The X166$\alpha$ line has a different spatial distribution compared to other
RRLs observed toward NGC 2024 and is probably associated with
cold dust clouds. Based on the expected low depletion of sulphur in such clouds and
the $-$8.1 \kms\ velocity separation between X166$\alpha$ and C166$\alpha$
lines, we interpret that the X166$\alpha$ transition arises from sulphur.
\end{abstract}

\section{Introduction}
\label{sec:intro}

NGC 2024 (Orion B, W12, Flame Nebula) is a well studied starforming region
located at a distance of 415 pc
(\nocite{a82}Anthony-Twarog 1982) in an extended ($\sim$ 1\ddeg.5 $\times$ 4\ddeg) molecular
cloud complex of mass
$\sim$ 10$^5$ \Msun (\nocite{tetal73}Tucker \etal 1973). It is located about 4\ddeg\ north of
the Great Orion Nebula (Orion A). The optical emission from the nebula
is obscured by a dark dusty structure in the molecular material. 
The main ionizing star is IRS2b, a star in the spectral range O8 V -- B2 V 
(\nocite{betal03}Bik \etal 2003). 

At radio
wavelengths, NGC 2024 is a source of abundant molecular
and recombination line emission (see for example, \nocite{betal10}Buckle \etal 2010,
\nocite{aetal90}Anantharamaiah \etal 1990).
Radio recombination line (RRL) emission from NGC 2024 is particularly striking
as spectral transitions of hydrogen, carbon and a heavy element are observed.
The richness of RRL emission toward NGC 2024 has led to several single dish recombination
line observations (\nocite{petal67}Palmer \etal 1967, \nocite{g69}Gordon 1969,
\nocite{c73}Chaisson 1973, 
\nocite{metal75}MacLeod \etal 1975, \nocite{retal77}Rickard \etal 1977,
\nocite{wt75}Wilson \& Thomasson 1975, \nocite{petal77}Pankonin \etal 1977, 
\nocite{ketal82}Kr\"{u}gel \etal 1982).

The hydrogen radio recombination lines (HRRLs) observed at frequencies \lsim 2 GHz
show broad ($\sim$ 30 \kms) and narrow (\lsim 4 \kms) components. 
The broad HRRL arises from the \HII\ region and is 
also observed at frequencies $>$ 2 GHz (\nocite{ketal82}Kr\"{u}gel \etal 1982). 
The central velocity and turbulent width features of broad HRRL as well as
the radio continuum emission suggest that the \HII\ region is ionization
bounded to the S (\nocite{ketal82}Kr\"{u}gel \etal 1982). 
The narrow hydrogen line, referred to as H$^0$RRL, 
has an LSR velocity $\sim$ 9 \kms\ (\nocite{betal70}Ball \etal 1970). 

The carbon radio recombination line (CRRL) observed toward NGC 2024 at low frequencies 
(\lsim 2 GHz) has an LSR velocity of $\sim$ 9 \kms. At these frequencies, 
the CRRL emission is dominated by stimulated
emission, suggesting that the carbon line originates from partially
ionized gas in front of the \HII\ region (\nocite{d74}Dupree 1974, 
\nocite{hw74}Hoang-Binh \& Walmsley 1974,
\nocite{wt75}Wilson \& Thomasson 1975, \nocite{petal77}Pankonin \etal 1977, 
\nocite{ketal82}Kr\"{u}gel \etal 1982).  This partially ionized gas,
located at the interface between the \HII\ region and molecular
material, is referred to as photo-dissociation region (PDR). 
The high resolution (45\arcsec) interferometric observation of RRLs
toward NGC 2024 at 1.4 GHz shows velocity and spatial coincidence of
CRRL and H$^0$RRL emission, suggesting that
the two line forming regions coexist (\nocite{aetal90}Anantharamaiah \etal 1990).

The CRRLs observed at high frequencies (\gsim 5 GHz) have LSR velocity
$\sim$ 10 \kms\ (\nocite{petal77}Pankonin \etal 1977) 
similar to that of line emission from 
dense molecular material at the far side of the \HII\ region. This
similarity suggests that the high frequency carbon lines originate from PDR
located at the far side (\nocite{petal77}Pankonin \etal 1977,
\nocite{ketal82}Kr\"{u}gel \etal 1982).

The spectrum at frequencies \lsim 2 GHz
also shows a line component with LSR velocity $\sim$ 0 \kms\ with respect to 
carbon line. This line is attributed to emission from a heavy element and 
is referred to as Xn$\alpha$, where n is the principal quantum number. 

In addition to RRL detections, there are numerous observations of molecular
lines (CS, HCN, OH, H$_2$CO, CO; \nocite{betal10}Buckle \etal 2010,
\nocite{getal93}Graf \etal 1993 and reference therein),
IR fine structure lines (CI, CII; eg. \nocite{getal12}Graf \etal 2012 and reference therein),
\HI\ 21cm line (\nocite{vetal93}van der Werf \etal 1993 and reference therein) and
IR continuum (\nocite{metal92}Mezger \etal 1992 and reference therein) toward NGC 2024.
High angular resolution ($\sim$ 13\arcsec)
H$_2$CO observations at 6 cm have shown absorption lines against the \HII\ region
with a velocity of $\sim$ 9 \kms\ (\nocite{cetal86}Crutcher \etal 1986). 
This gas, located in front of the \HII\ region, has a density, $n_H$ $\sim$ 
8 $\times$ 10$^4$ \cmthree (\nocite{hww80}Henkel \etal 1980).
Millimeter wave emission lines of H$_2$CO, on the other hand, are observed at
a velocity $\sim$ 10.5 \kms\ with a derived gas density, $n_H$ $\sim $  3.6 $\times$ 10$^5$ \cmthree
(\nocite{mw93}Mangnum \& Wootten 1993). The study of CS line emission at 10.5 \kms\ has shown
that the core of the molecular cloud may  have density 
$\sim$ 2 $\times$ 10$^6$ \cmthree
(\nocite{setal84}Snell \etal 1984). The presence of a two component gas structure is also
evident from other molecular line (eg. CO) observations (eg. \nocite{getal93}Graf \etal 1993),
as well as \HI\ absorption line data (\nocite{vetal93}van der Werf \etal 1993).

The picture of the NGC 2024 gas distribution that emerges from previous studies
is the following : the low density ($\sim$ 10$^5$ \cmthree) component, part
of which is located in front of the \HII\ region, has a total mass $\sim$ 300 \Msun
and the dense ($\sim$ 10$^6$ \cmthree) gas, located at the far side and S of the \HII\ region, has
size 0.04 $\times$ 0.3 pc extending along a N-S direction with mass $\sim$ 140 \Msun
(\nocite{betal89}Barnes \etal 1989, \nocite{metal92}Mezger \etal 1992). In this picture, 
the PDR responsible for the low frequency CRRL and H$^0$RRL emissions is co-located 
with the foreground low density gas while the high frequency
CRRLs originate from the PDR associated with the denser material at the far side of
the \HII\ region (eg. \nocite{wt75}Wilson \& Thomasson 1975, 
\nocite{petal77}Pankonin \etal 1977, \nocite{ketal82}Kr\"{u}gel \etal 1982).

Almost all previous studies of RRL emission toward NGC 2024 were 
single dish observations. At frequencies \lsim 15 GHz, the angular
resolutions of these observations are several arcminutes. The
angular resolution of the single dish observations are, however, 
inadequate for modeling CRRL
and H$^0$RRL emission as the beam dilution factor is unknown. High angular 
resolution images at frequencies \lsim 15 GHz with full spatial
information are needed to better estimate the physical properties
of the PDR.  
In this paper, we present new high sensitive, high angular resolution 
RRL observations toward NGC 2024 at 1.4 GHz
(angular resolution $\sim$ 70\arcsec) 
and 4.8 GHz (angular resolution $\sim$ 17\arcsec)
with the VLA (Very Large Array). To obtain
the complete spatial information at 4.8 GHz, we combined RRL observations
made with the GBT (Green Bank Telescope) to VLA data. The observations and data analysis
methods are described in Section~\ref{sec:obs}. The observed features
of the continuum and RRL emission are discussed in Section~\ref{sec:cont}
and \ref{sec:line}, respectively. The data set is used
to model the physical properties of the region responsible for CRRL and
H$^0$RRL emission, which is discussed in Section~\ref{sec:model}.
In Section~\ref{sec:mag}, the physical
properties are used to derive the magnetic field in the CRRL emitting
region using method proposed by \nocite{r07}Roshi (2007). Discussion
and summary of the results are given in Section~\ref{sec:sum}.

\section{Observation and data reduction}
\label{sec:obs}

We observed recombination lines toward NGC 2024 using the D-configuration of the VLA
near 1.4 and 4.8 GHz in dual polarization mode.
The RRLs observed are the 166$\alpha$ transition of carbon (1425.444 MHz)
and hydrogen (1424.734 MHz) near 1.4 GHz and the 110$\alpha$ transition
of carbon (4876.589 MHz) and hydrogen (4874.157 MHz) near 4.8 GHz.
The uv sampling of VLA at the shortest spacing is not adequate to image the
full angular extent of NGC 2024 at 4.8 GHz. Therefore we observed the source
in ``On-the-fly'' mapping mode with the Green Bank Telescope (GBT)
at 4.8 GHz and combined this data with that obtained with
the VLA at the same frequency. Both VLA and GBT observations were 
done during test times.
A summary of the observation log is given in Table~\ref{tab1}.

The interferometric data was analyzed using Astronomical Image
Processing Software (AIPS).  After editing and calibration of the default continuum
data provided by the VLA (channel 0 data),
the flag and calibration tables were transferred to the spectral line
data set.  The system band-shapes were obtained with the AIPS task BPASS.
3C48 was observed at both frequencies for bandpass
and flux density calibration. For continuum subtraction 
the task UVLSF was used, which also provided the continuum data. 
The spectral cubes and continuum images were obtained with the 
AIPS task IMAGR.
The final images were corrected for primary beam gain variation
using the task PBCOR.

An ``On-the-fly'' map of size about 0\ddeg.5 $\times$ 0\ddeg.5 centered at
RA(2000) $05^h41^m43^s.1$, DEC(2000) $-01^{o}54^{'}00^{''}$ was made with the
GBT. The telescope was scanned along RA
and at the end of every 6$^{th}$ RA scan a reference
position (RA(2000) $05^h35^m04^s.2$, DEC(2000) $-01^{o}54^{'}38^{''}$) was
observed. The FWHM beam width of the GBT at the observing freq of
4.9 GHz is 2\arcmin.5. The RA scan rate and integration time
were chosen to provide about 4 samples within the 2\arcmin.5 beam.
Along the declination direction, the telescope was scanned at an interval of 0\arcmin.5.
The GBT data were analyzed using GBTIDL. A modified version of
task `getrefscan' was used to obtain the spectra in units of
antenna temperature and re-sampled to the spectral resolution corresponding
to the VLA data. A telescope gain of 2 K/Jy was used to convert
the spectra in units of Jy per beam.
The telescope gain was also measured by observing the calibrator
source 3C123. A 3$^{rd}$ order polynomial baseline was removed
from each of the calibrated spectrum. The zeroth order coefficient
of the polynomial baseline was used to obtain the continuum data set.
The calibrated data were converted to a UVFITS file
using the program SDFITS2UV, developed by Glenn Langston.
The UVFITS file was transferred to AIPS and the image was made
using the SDGRID task. The VLA and GBT data were combined using 
the CASA function Feather.

\section{Continuum Emission}
\label{sec:cont}

\begin{figure}
\plotone{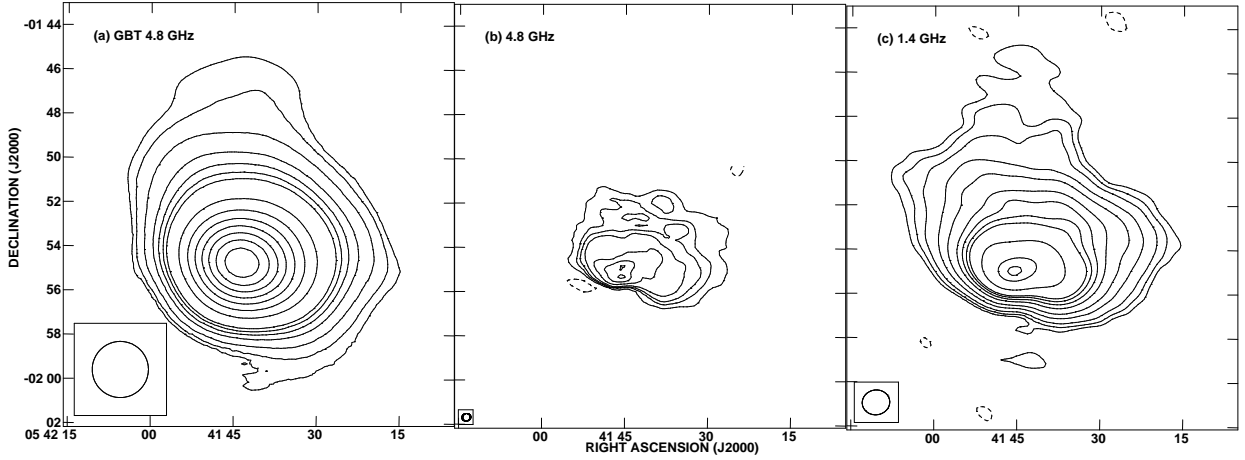}
\caption{
The continuum images of NGC 2024. (a) The image made with the GBT at 4.8 GHz. The
angular resolution of the image is 152\arcsec. The contour
levels are ($-$3,3,5,10,20,30,40,50,100,150,200,250,300,350,400) $\times$ 50 mJy/beam.
(b) The image made from the combined GBT and VLA data at 4.8 GHz with an angular
resolution of 22\arcsec $\times$  20\arcsec ($-$71$^{o}$). The contour levels
are ($-$1,1,2,3,5,10,20,30,40,50,100,150,200) $\times$ 50 mJy/beam. (c) VLA image
of NGC 2024 at 1.4 GHz. The angular resolution of the image is 75\arcsec $\times$  
67\arcsec ($-$81$^o$). The contour levels
are ($-$1,1,2,4,10,20,40,60,80,100,200,400,500) $\times$ 20 mJy/beam.  
\label{fig1} } 
\end{figure}

The continuum images of NGC 2024 at 1.4 GHz and 4.8 GHz are shown in Fig.~\ref{fig1}.
The angular resolution of the 1.4 GHz image obtained from VLA data is
75\arcsec $\times$  67\arcsec\ and the total flux density is
61.0 $\pm$ 1.2 Jy. The image at 4.8 GHz, obtained from the combined GBT and VLA data, has
an angular resolution of 22\arcsec $\times$  20\arcsec. The total flux density in this
image is 52.5 $\pm$ 0.2 Jy. This value is less than the flux-density (57 $\pm$ 3 Jy) 
in the low resolution (152\arcsec $\times$ 152\arcsec) GBT image.
The half power beamwidth of the VLA antenna at 4.8 GHz is $\sim$ 10\arcmin\ and,
for our observations, centered at RA(2000)  $05^h41^m44.5^s$, 
DEC(2000) $-01^{o}54^{'}39^{''}$. As seen in Fig.~\ref{fig1}, the continuum
emission in the GBT image and the 1.4 GHz VLA image extends beyond the
10\arcmin\ beamwidth of the VLA antennas at 4.8 GHz.  We therefore attribute the difference 
in flux density in the 4.8 GHz GBT image and the 4.8 GHz combined data image to 
emission extending beyond the half power beam width of the VLA antenna. 
The flux density obtained for NGC 2024 at the two frequencies are consistent with earlier
observations (eg. \nocite{betal89}Barnes \etal 1989).

Our continuum images are consistent with higher resolution 
(4\arcsec.9 $\times$ 3\arcsec.8) images of \nocite{betal89}Barnes \etal (1989)
as well as those presented by \nocite{aetal90}Anantharamaiah \etal (1990).
A radio plume extending
roughly 5\arcmin\ northwards from the main nebula emission is detected in
both the 1.4 GHz VLA image and 4.8 GHz GBT image (see Fig.~\ref{fig1}).
The radio plume was detected and imaged earlier at other frequencies as well
(see, for example, \nocite{setal97}Subrahmanyan \etal 1997 at 327 MHz). 

\section{Recombination Line Emission towards NGC 2024}
\label{sec:line}

We have detected broad ($\Delta V \sim 25$ \kms) H166$\alpha$
and H110$\alpha$ lines at the two observed frequencies (1.4 and 4.8 GHz). 
The C166$\alpha$ and 
C110$\alpha$ transitions are also detected. In addition, the H$^0$166$\alpha$
and X166$\alpha$ lines are detected at 1.4 GHz observations. 
In this Section, we present images of the line emission and their observed
characteristics.

\subsection{H and H$^0$  Recombination Lines}

\begin{figure}
\plotone{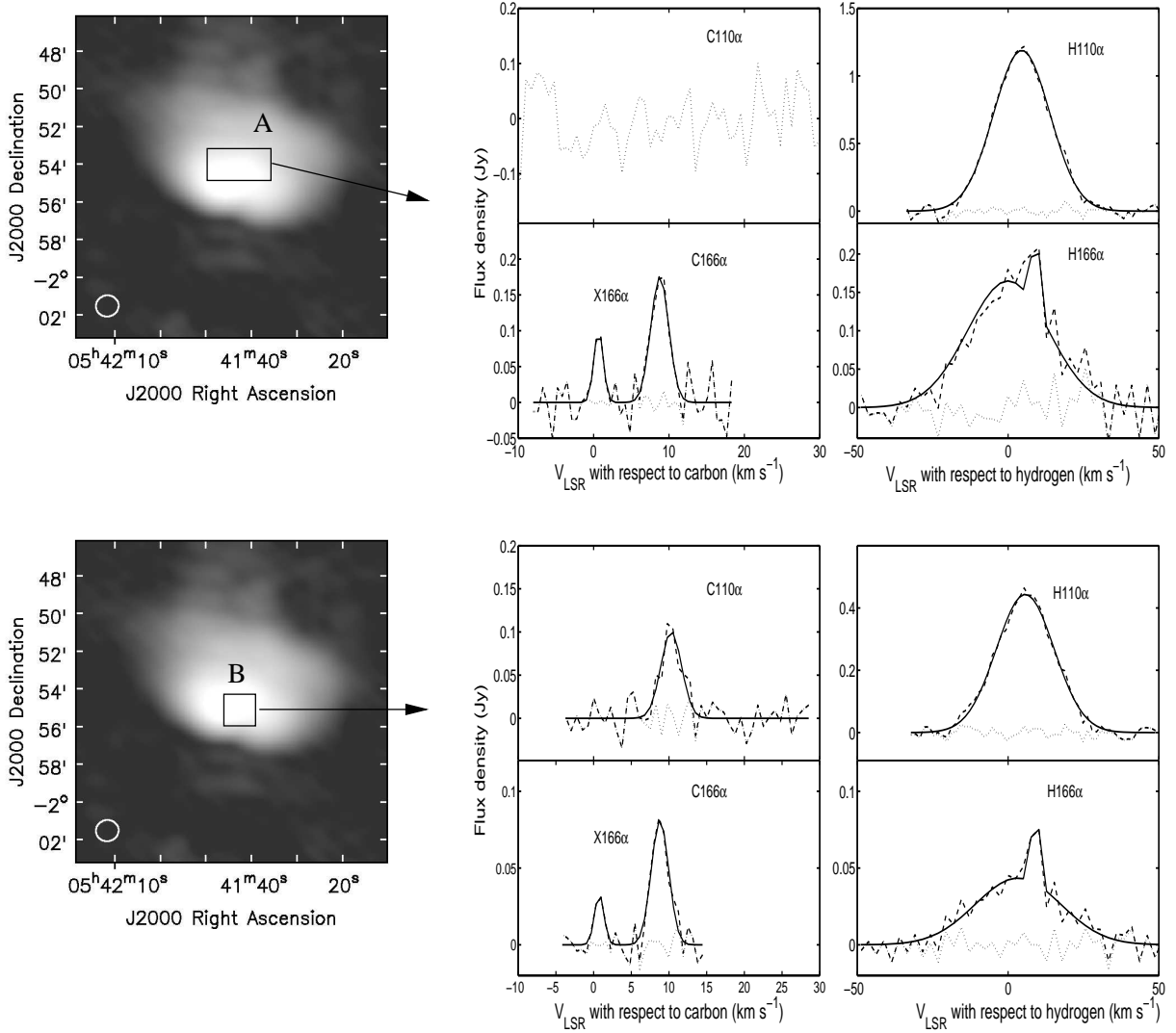}
\caption{
Representative spectra of the observed lines toward NGC 2024. The spectra are obtained by
averaging the data over the region A (top) and B (bottom). The
observed transitions are marked on the plot. The observed spectrum is shown by dashed line,
Gaussian component model for the line emission is shown by the solid line and
the residual obtained after subtracting the
model from the observed spectrum is shown by the dotted line. 
\label{fig2} } 
\end{figure}

\begin{figure}
\includegraphics[height=6.4in, width=6in, angle=0]{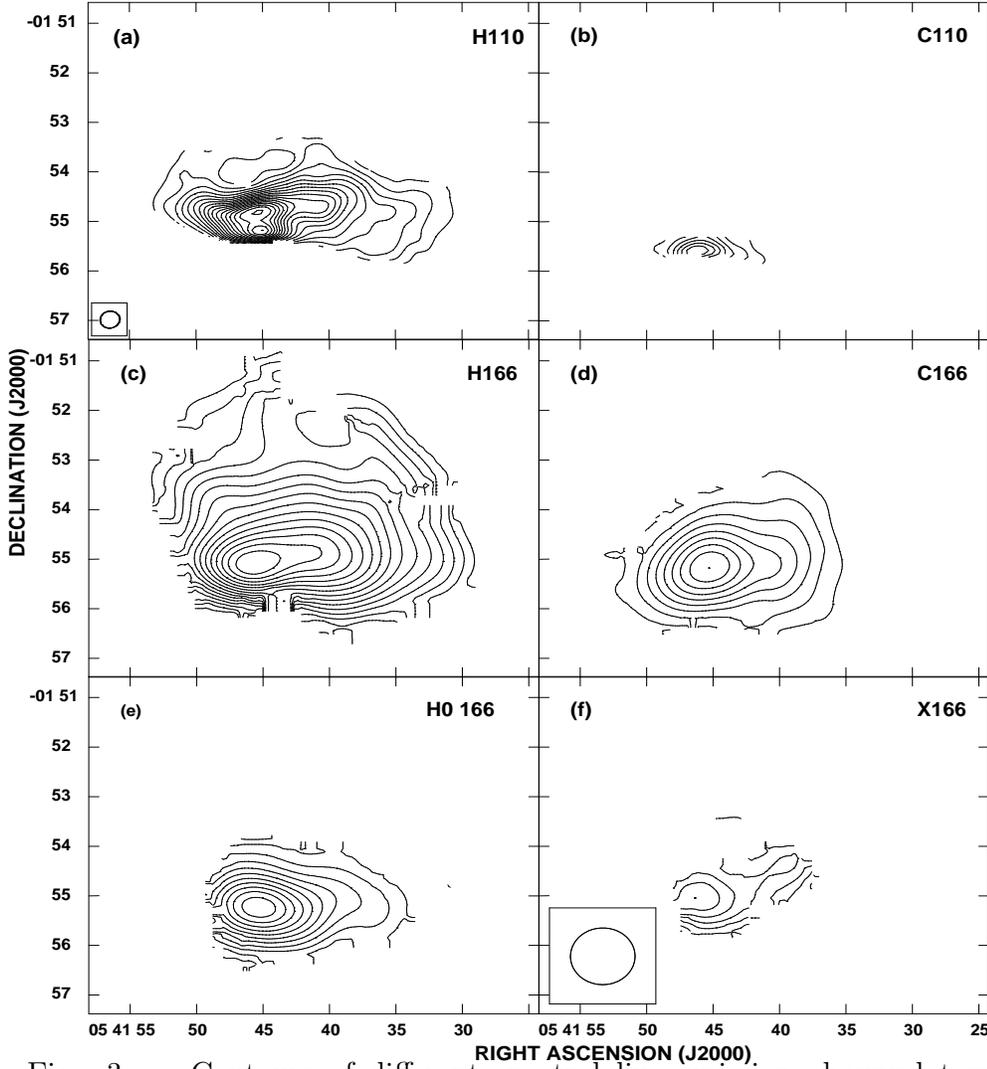}
\caption{
Contours of different spectral line emission observed toward NGC 2024. 
(a) Contours of the peak line amplitude of the H110$\alpha$ emission.
The contour levels are (1 to 25) 
$\times$ 5.2 mJy/beam. The angular resolution of the line image is 
22\arcsec $\times$  20\arcsec ($-$71$^{o}$).
(b) Contours of the peak line amplitude of the C110$\alpha$ emission.
The contour levels are (1 to 20)
 $\times$ 5.1 mJy/beam. The angular resolution of the image is 
22\arcsec $\times$  20\arcsec ($-$70$^{o}$).
(c) Contours of the peak line amplitude of the H166$\alpha$ emission.
The contour levels are (5 to 25) $\times$ 3.2 mJy/beam. 
The angular resolution of the image is 75\arcsec $\times$  
67\arcsec ($-$82$^{o}$). 
(d) Contours of the peak line amplitude of the C166$\alpha$ emission.
The contour levels are (1 to 20) $\times$ 13.2 mJy/beam. 
The angular resolution of the image is 73\arcsec $\times$  
69\arcsec ($-$89$^{o}$). 
(e) Contours of the peak line amplitude of the H$^0$166$\alpha$ emission. 
The contour levels are (1 to 20 ) $\times$ 5.3 mJy/beam. 
The angular resolution of the image is 75\arcsec $\times$  
67\arcsec ($-$82$^{o}$). 
(f) Contours of the peak line amplitude of the X166$\alpha$ emission.
The contour levels are (1 to 20) $\times$ 5.3 mJy/beam. 
The angular resolution of the line image is 73\arcsec $\times$  
69\arcsec ($-$89$^{o}$) 
\label{fig3} } 
\end{figure}

Representative profiles of H110$\alpha$ and H166$\alpha$ line emission
are shown in Fig.~\ref{fig2}. The line parameters are summarized in Table~\ref{tab2}.
The regions over which the data is averaged to obtain the profiles
are also included in Table~\ref{tab2}.
The H166$\alpha$ line emission shows a narrow component
($\Delta V \sim$ 1.7 \kms) in addition to a broad feature 
($\Delta V \sim$ 33 \kms). At
4.8 GHz only the broad component is detected, with a  line width
$\sim 23$ \kms. The spatial distributions of the broad HRRL at the
two observed frequencies are shown in Fig.~\ref{fig3}. 

The line-to-continuum ratio of the broad H110$\alpha$ provides an
electron temperature of 7200 K (assuming LTE). This value is consistent
with the electron temperature
obtained by \nocite{ketal82}Kr\"{u}gel \etal (1982) from H76$\alpha$
observations. As mentioned above, the observed H166$\alpha$ line width 
is 1.5 times higher than the width of H110$\alpha$ line. We suggest
that the increase in line width is due to pressure broadening. The
inferred local electron density is $\sim$ 1400 \cmthree. The transverse size
of the \HII\ region is $\sim$ 0.3 pc.  If we assume that the
extent of the \HII\ region along the line-of-sight is also $\sim$ 0.3 pc,
then the derived electron density indicates an emission measure of 
6 $\times$ 10$^5$ \cmsix pc.
This value is comparable to the emission measure derived from other
observations (eg. \nocite{setal97}Subrahmanyan \etal 1997 at 327 MHz).

The H$^0$166$\alpha$ line is detected over a $\sim$ 2\arcmin $\times$
4\arcmin\ region and is centered near the continuum peak (see Fig.~\ref{fig3}). 
The spatial distribution of line emission
obtained from our data is similar to that observed earlier by 
\nocite{aetal90}Anantharamaiah \etal (1990).
The LSR velocity ($\sim$ 9 \kms)
of this feature is about 5 \kms\ higher than that of the broad
H166$\alpha$ line and is  closer to the central velocity of the 
C166$\alpha$ line emission (see below).
The spatial distributions of both H$^0$166$\alpha$ and C166$\alpha$ 
lines are also in good agreement.

\subsection{Carbon Recombination Lines}
\label{sec:cline1}

Representative profiles and images of carbon line emission
are shown in Fig.~\ref{fig2} and Fig.~\ref{fig3}, respectively.
The C110$\alpha$ line emission is detected over a region $\sim$ 0\arcmin.5
$\times$ 2\arcmin.5, confined to the southern ridge along the
\HII\ region -- molecular cloud interface. 
The integrated line flux density of C110$\alpha$ transition
is $\sim$ 425 mJy. The central velocity
of the C110$\alpha$ line ranges between 10.0 and 10.5 \kms , with an
average 10.25 \kms. 

The angular size of C166$\alpha$ line emission is $\sim$ 2.5\arcmin
$\times$ 5\arcmin, larger than the angular size of
C110$\alpha$ line emission. The distinct feature of C166$\alpha$ emission
is that the LSR velocity ranges from 8.7 \kms\ to 10.5 \kms.
The C166$\alpha$ line emission to the N 
has velocity $\sim$ 8.8 \kms\ while the emission to the SE, nearer the
ionization front, has a velocity $\sim$ 10.0 \kms. The
similar central velocity of the southern C166$\alpha$ emission 
and C110$\alpha$ line emission suggests that the two line
forming regions are associated. 

The spatial distributions of the H$^0$166$\alpha$ and C166$\alpha$
line emissions in the VLA images are similar (see Fig.~\ref{fig3}). Further,
the central velocity of H$^0$166$\alpha$ line ($\sim$ 9 \kms\ ) is 
comparable to the northern C166$\alpha$ line emission.
The line widths of the two lines are also comparable.
These similarities suggest
that the two line forming regions are spatially co-located
(see also \nocite{aetal90}Anantharamaiah \etal 1990).  

Molecular line observations over a wide range of frequencies (1.6 GHz to 100 GHz)
have shown that the \HII\ region is partially enclosed by the parent molecular cloud
(\nocite{betal89}Barnes \etal 1989; see also Fig.~\ref{fig5} and Section~\ref{sec:sum}).
The LSR velocity of the gas in front
of the \HII\ region is $\sim$ 9 \kms , similar to the central
velocity of the northern C166$\alpha$ line emission. This similarity in LSR velocity
suggests that the low frequency ($<$ 2 GHz) carbon line forming region
is associated with  molecular gas in front of the \HII\ region.
The dense gas located on the far
side has an LSR velocity 10.5 \kms\ 
(see for example \nocite{mw93}Mangnum \& Wootten 1993).
This LSR velocity compares
well with that of the C110$\alpha$ line, suggesting that the line
emission is associated with the dense gas located on the far side of the \HII\ region.
(see also Section~\ref{sec:sum} and Fig.~\ref{fig5}).

\subsection{X Recombination Lines}

The velocity of X166$\alpha$ is $-$8.1 \kms\ relative to the C166$\alpha$ line,
which implies a transition from a heavy element such as S, Mg, Si or Fe.
The spatial distribution of X166$\alpha$ emission is shown 
in Fig.~\ref{fig3} and representative line profiles 
are shown in Fig~\ref{fig2}. 
The spatial distribution and line width of X166$\alpha$ are different 
from those of C166$\alpha$ and H$^0$166$\alpha$. 
We conclude that the X166$\alpha$ line emission
is not associated with the carbon and H$^0$ line forming regions.
\nocite{aetal90}Anantharamaiah \etal (1990) came to a similar conclusion
based on their 1.4 GHz observations.
The X166$\alpha$ line width is about 0.8 \kms, which is about one third 
that of the carbon line. This smaller
width may suggest that the X166$\alpha$ line is associated with colder ($<$ 100 K)
dust clouds. In such clouds sulphur is less depleted relative
to other heaver elements such as Mg, Si and Fe 
(\nocite{petal77}Pankonin \etal 1977). 
Based on the expected low depletion of sulphur
and velocity separation, we suggest that the X166$\alpha$ 
line emission arises from sulphur.

\section{Models for Carbon and H$^0$166$\alpha$ Line emission}
\label{sec:model}

We consider a simplified model consisting of homogeneous `slabs' of
PDR, co-located with the molecular gas. 
The recombination line flux density from the PDRs is obtained
by solving the radiative transfer equation for non-LTE cases.
A program to compute the non-LTE departure
coefficients $b_n$ and $\beta_n$ was  developed. This program
uses new numerical techniques to solve for the departure coefficients
based on the formulism
developed by \nocite{bs77}Brocklehurst \&  Salem (1977) and
\nocite{ww82}Walmsley \& Watson (1982).
For a given gas temperature ($T_{PDR}$) and electron density
($n_e^{PDR}$), we solve separately for the departure co-efficients of 
hydrogenic and carbon cases. 
The atomic level population is altered by stimulated emission,
which is due to background radiation field. 
Therefore, a background thermal radiation field from the
\HII\ region (see below) is included when solving for the departure
co-efficients. The depletion of carbon in PDR is inferred to
be small ($\sim$ 25 \% \nocite{nwt94}Natta \etal 1994).
Therefore in our computation the abundance of carbon is taken to be 0.75 of the cosmic value
3.98 $\times$ 10$^{-4}$ (\nocite{m74}Morton 1974).
The dielectronic like recombination process that
modifies the level population of the carbon atom (\nocite{ww82}Walmsley \& Watson 1982)
is also included in the calculation.
The coefficients are computed by considering
a 10000 level atom with the boundary condition
$b_n \rightarrow 1$ at higher quantum states.

The line brightness temperature, $T_{LB}$, due to the slab 
is given by (\nocite{s75}Shaver 1975)
\begin{equation}
T_{LB}  =  T_{bg,\nu} + T_{in,\nu}\ ,
\label{eq:line1}
\end{equation}
where
\begin{eqnarray}
T_{bg,\nu} & = & T_{0bg,\nu} e^{-\tau_{C\nu}}(e^{-\tau_{L\nu}} - 1) \ ,\nonumber \\
T_{in,\nu} & = & T_{PDR}\left( \frac{b_m \tau_{L\nu}^* + \tau_{C\nu}}{\tau_{L\nu} + \tau_{C\nu}}
          (1 - e^{-(\tau_{L\nu} + \tau_{C\nu})}) - (1 - e^{-\tau_{C\nu}})\right).
\label{eq:line}
\end{eqnarray}
$T_{bg,\nu}$ is the contribution to the line temperature due to the
background radiation field and $T_{in,\nu}$ is the intrinsic line emission from
the slab. In Eq.~\ref{eq:line},  $T_{0bg,\nu}$ is the background
radiation temperature due to the \HII\ region;
$\tau_{C\nu}$ is the continuum optical depth of the PDR. The non-LTE line
optical depth of the spectral transition from energy state $m$ to $n$,
$\tau_{L\nu}$, is defined as $\tau_{L\nu}  =  b_n \beta_n \tau_{L\nu}^*$
where $\tau_{L\nu}^*$ is the LTE line optical depth,
$b_n$ and $\beta_n$ are the departure coefficients of energy state $n$.
$\tau_{L\nu}^*\; \propto\;  n_e^{PDR} n_{i} l$, where $n_{i}$ is the
ion density and $l$ is the line-of-sight path length of the PDR;
$n_{i}$  is equal to  $n_{C^+}$
for the carbon line computation, while  $n_{i}$ is equal to  $n_{H^+}$
for the hydrogen line
computation. For the carbon line computation we assume
that $n_e^{PDR} = n_{C^+} = n_e$; thus  $\tau_{L\nu}^*\; \propto\;  n_e^2 l$.
The neutral density, $n_H$, in the PDR
is obtained as $n_H = n_e^{PDR} / (0.75 \times 3.98 \times 10^{-4})$.
The line brightness temperature is
finally converted to flux density using the observed angular sizes
of the line emission, tabulated in Table~\ref{tab2}.

\subsection{Foreground Carbon line emission}

The observed parameters of the line emission from the foreground PDR is obtained
by averaging the spectra over region A shown in Fig.~\ref{fig2}. 
The region is selected such that it is northward of the C110$\alpha$
line emission, thus avoiding contribution from the PDR emission from
the far side of the \HII\ region.
The line and continuum parameters are given in Table~\ref{tab2}. 
The images at the two observed frequencies are convolved to 
the same angular resolution (73\arcsec $\times$  69\arcsec) before 
obtaining these parameters.  The
continuum flux density at 1.4 and 4.8 GHz toward region A
is consistent with emission from an \HII\ region with parameters Te $\sim$ 7000 K and
EM $\sim$ 3.5 $\times$ 10$^5$ pc \cmsix. 

\begin{figure}
\includegraphics[height=5.5in, width=6.0in, angle=0]{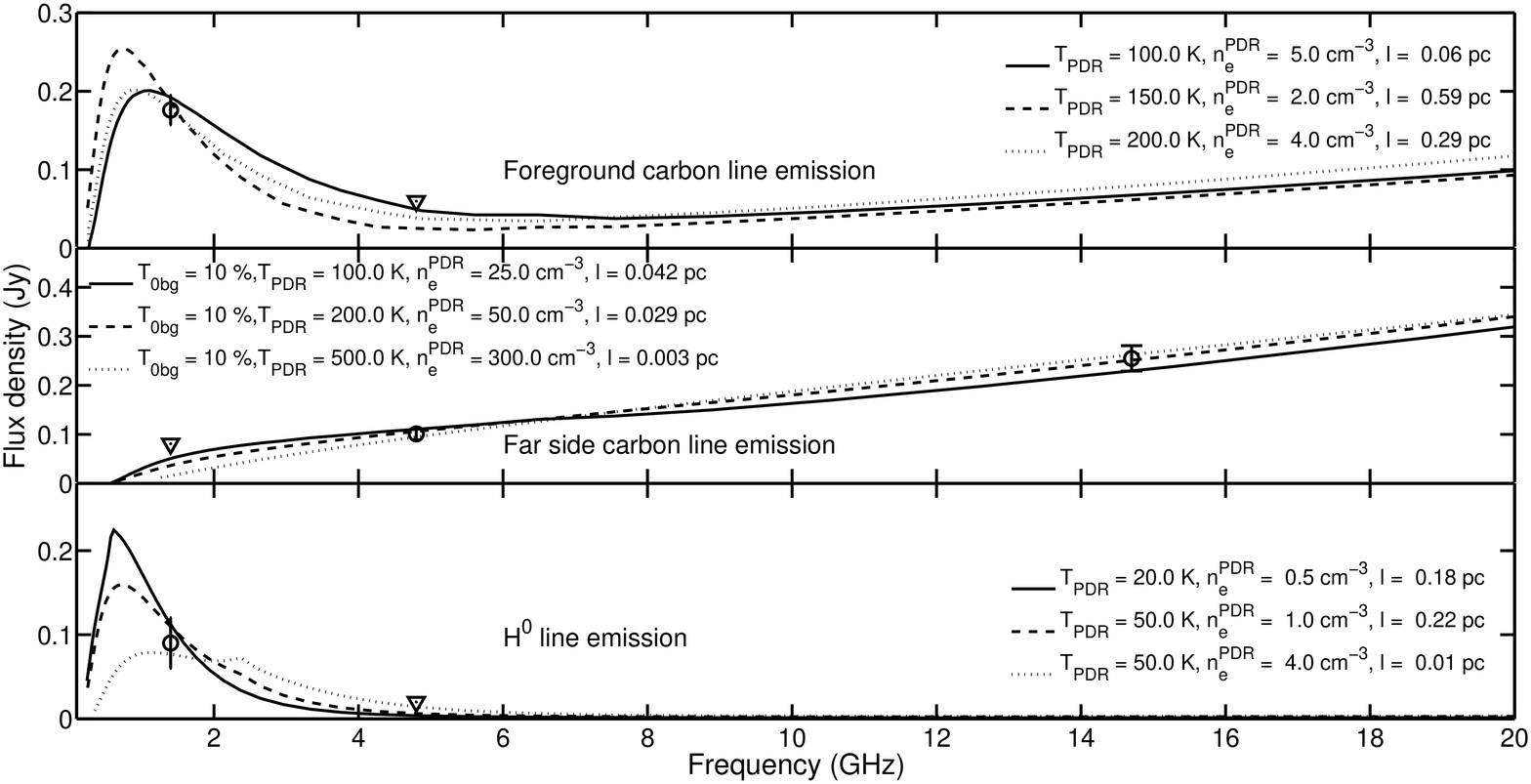}
\caption{Line flux density as a function of frequency for 
a subset of models consistent with RRL observations. 
Results of modeling for the foreground PDR, far side PDR  
and H$^0$ region are shown respectively on 
the top, middle and bottom panels. The flux density of the detected 
lines along with $\pm$ 1$\sigma$ error bars are shown in the top and
middle panel; the error bar is $\pm$ 3$\sigma$ in the bottom panel. 
The upper limit for line flux density at 4.8 GHz are shown by 
triangles in the top and bottom panels. The flux density of
the C166$\alpha$ line is shown as an upper limit in the middle panel,
since this line could have a contribution from the foreground PDR. 
The line flux density of C76$\alpha$ transition (near 15 GHz) shown
in the middle panel is estimated from Kr\"{u}gel \etal (1982) data.
The model parameters 
corresponding to the different curves are shown on each plot.  
\label{fig4} } 
\end{figure}

The foreground PDR
is approximated with a slab of partially ionized gas with background
radio continuum emission due to \HII\ region with the above mentioned 
properties.  Departure coefficients are
obtained for a set of $T_{PDR}$ between 100 and 200 K and
$n_e^{PDR}$ between 1 and 10 \cmthree. The model flux density
for these gas properties are required to be consistent with the observed
values. This requirement is realized
by varying $l$. A range of models are  found to be
consistent with the observed RRL data. Model flux density vs 
frequency for three gas properties are shown
in Fig.~\ref{fig4}. Modeling shows that the upper limit on $n_e^{PDR}$
is $\sim$ 5 \cmthree, corresponding to a neutral
density of 1.7 $\times 10^4$ \cmthree. For this density, the
thickness of the PDR would be 0.06 pc, if the gas temperature 
is 100 K. To further constrain the PDR properties,
we compare the neutral density with values obtained for the
foreground gas from molecular line observations. The density
obtained from H$_2$CO absorption studies is $\sim$ 4 $\times$ 10$^4$ \cmthree.
This density is in rough agreement with the neutral density
for the $n_e^{PDR} \sim 5$ \cmthree model. Based on this agreement, we conclude
that the physical properties of the foreground PDR is
$T_{PDR} \sim 100$ K, $n_e^{PDR} \sim 5$ \cmthree,
$n_H \sim 1.7 \times 10^4$ \cmthree and $l \sim 0.06$ pc.

The \HII\ region in NGC 2024 is partially enclosed in the parent molecular
cloud (see Fig.~\ref{fig5}).
In this geometry, a PDR with {\em similar physical properties
as that of the foreground PDR} should exist on the far side of the \HII\ region.
This PDR is referred to as `far side low-density PDR' in Fig.~\ref{fig5}.
We investigate whether such a PDR would produce observable RRLs. 
For this investigation,
the line flux density is calculated by setting the
background continuum emission to zero (ie $T_{0bg} = 0$ in Eq.~\ref{eq:line}).
The computed line flux density is \lsim 10 mJy at all 
the observed frequencies and therefore we conclude that RRLs from such a PDR
would not be detectable. This lower flux density is
due to lack of stimulated emission in the absence of background
continuum emission.  

\subsection{Carbon line emission from the far side of the \HII\ region}

The observed parameters of the carbon line from the far side of the \HII\ region
are obtained by averaging the data over region B shown in Fig~\ref{fig2}.
Region B is chosen so as to minimize contamination
from the foreground PDR at 1.4 GHz.
The images at the two observed frequencies are convolved to 
the same angular resolution (73\arcsec $\times$  69\arcsec) to 
obtain these parameters.  
The line and continuum parameters are listed in Table~\ref{tab2}.
The continuum emission within region B is consistent with
emission from ionized gas of Te $\sim$ 7000 K, EM $\sim$ 5 $\times$ 10$^5$ pc \cmsix.

The LSR velocity of carbon line (10.3 \kms ),
is similar to the velocity of dense molecular tracers observed
from the far side of the \HII\ region.
\HI\ , OH and H$_2$CO lines show absorption features near
10 \kms\ as well as a component near 9 \kms\ . The component
near 9 \kms\ is due to foreground gas.
The absorption line near 10 \kms\ indicates that a
part of the continuum emission is located behind the
gas responsible for this line component
(see also \nocite{cetal99}Crutcher \etal 1999).
As mentioned earlier, the PDR responsible for the
10 \kms\ carbon line emission is co-located with this gas.
Thus the PDR responsible for
the 10 \kms\ line component must be located {\em within}
the \HII\ region (see Fig.~\ref{fig5} and Section~\ref{sec:sum}). 
As discussed below such a geometry
is required to explain the CRRL emission from this PDR.

To model carbon line emission, we approximate the far 
side PDR to be a single slab of partially ionized gas.
OH observation shows that a small part of the radio
continuum is located behind this PDR (\nocite{cetal99}Crutcher \etal 1999).
For modeling, we assume that $\sim$ 10 \% of the observed 
radio continuum emission is located behind the PDR.
The line flux densities used to constrain the 
model are those of C166$\alpha$, C110$\alpha$ (see Table~\ref{tab2}) 
and C76$\alpha$ emission. The line flux density of C76$\alpha$ averaged over
region B is estimated to be 255$\pm$26 mJy from the
\nocite{ketal82}Kr\"{u}gel \etal (1982) observations. The 
foreground PDR contributes to the C166$\alpha$ emission, thus this 
line flux density defines an upper limit to the emission
from the far side PDR at 1.4 GHz. The departure coefficients
for $T_{PDR}$ between 100 and 500 K are computed 
with a thermal radiation field from ionized gas of  Te $\sim$ 7000 K, 
EM $\sim$ 5 $\times$ 10$^5$ pc \cmsix. Results from three representative 
models are shown in Fig.~\ref{fig4}. We found that for
gas temperature between 100 and 500 K, models with
$n_e^{PDR}$ in the range 25 to 300 \cmthree\ are consistent
with the carbon line observations.
The temperature of the PDR cannot be constrained.
Representative physical properties for the far side PDR
are $T_{PDR} \sim 200$ K, $n_e^{PDR} \sim 50$ \cmthree, $l \sim 0.03$ pc.
The C166$\alpha$ flux density predicted by the model is about half 
the observed value.
Since the electron density is an order of magnitude larger than that of the
foreground PDR, we refer to this interface region as the `far side dense PDR' (see Fig.~\ref{fig5}).
The neutral density in the PDR is 1.7 $\times 10^5$ \cmthree, 
comparable to the density derived from  H$_2$CO emission line (LSR velocity 10.5 \kms)
observations (3.6 $\times$ 10$^5$ \cmthree ; \nocite{mw93}Mangnum \& Wootten 1993).

\subsection{H$^0$ line forming region}
\label{sech0model}

For modeling H$^0$ line emission, the line parameters are obtained by averaging
the spectra over region B shown in Fig.~\ref{fig2}. The line
and continuum parameters are given in Table~\ref{tab2}.
The observed continuum emission from this region is
consistent with emission from ionized gas with $T_e \sim 7000$ K
and EM $\sim 5 \times 10^5$ pc \cmsix.
The upper limit on the temperature of the PDR obtained
from the width of H$^0$166$\alpha$ line is $\sim$ 70 K.
Therefore, for modeling H$^0$ line emission, we considered
$T_{PDR}$ in the range 20 to 70 K. The departure coefficients
for these models were obtained for the hydrogenic case.
Model flux density vs frequency is shown for 
three models in Fig.~\ref{fig4}.
Our data provide an upper limit for $n_e^{PDR}$ of $\sim$ 4 \cmthree
for the H$^0$ region; for higher density the model
line flux density exceeds the observed upper limit at 4.8 GHz. 
No additional constraints on the
PDR properties can be obtained from the existing data set.
Representative PDR properties are :  
$T_{PDR} \sim 50 $ K, $n_e \sim 4$ \cmthree and $l \sim 0.01$ pc.
As suggested in Section~\ref{sec:cline1}, 
the H$^0$ region is co-located with the foreground
carbon line forming region. 
If we assume that the neutral density of
the H$^0$ line forming region is similar to that of the
foreground PDR, the ionization fraction is $\sim$ 10$^{-4}$.

\section{Magnetic field in the PDR}
\label{sec:mag}

Alf$\acute{v}$en waves
in the PDR are strongly coupled to ions,
affecting the non-thermal width of the spectral lines.
\nocite{r07}Roshi (2007) used this effect to derive magnetic fields from the observed widths of
carbon recombination lines. The thermal contribution to the line width
is removed using the values for $T_{PDR}$ to obtain the non-thermal
width $V_{nth}$.  The magnetic field is
then obtained as:
\begin{equation}   
B  = \frac{\sqrt{3}}{2} \times \frac{V_{nth}}{\sqrt{8 ln(2)}} \sqrt{4\pi\rho},
\end{equation}
where $V_{nth}$ is in cm sec$^{-1}$, $\rho$ is the mass density
in grams and B is in Gauss. The density, $\rho$, is obtained from the neutral density in the
PDR by considering the effective mass of the gas as 1.4. 
The derived magnetic field for the foreground PDR is 60  $\mu$G 
and that for the far side dense PDR is 220 $\mu$G. The 
error in the estimated field strength is due to 
error in the observed line width and PDR modeling limitations.
We estimate the 
error in the derived field strength to be about 60 \%.  

\nocite{cetal99}Crutcher \etal (1999) used OH Zeeman observations
to determine the distribution  of magnetic field across NGC 2024. 
The angular resolution of their observation is $\sim$ 1\arcmin,
comparable to our observations.  The OH line component
used to derive the magnetic field has LSR velocity 10.2 \kms, similar
to the carbon line emission from the far side PDR. 
A peak field strength of 87 $\pm$ 5.5 $\mu$G is obtained
toward RA(2000): $05^h41^m41.5^s$, DEC(2000): $-01^{o}55^{'}04^{''}$.
Zeeman observations measure the line-of-sight component of the
magnetic field; the measured field strengths are scaled
by a factor of 2 to obtain the total magnetic field strength 
(\nocite{c99}Crutcher 1999). Our analysis provides the total 
magnetic field (see \nocite{r07}Roshi 2007) and the field
strength obtained toward RA(2000): $05^h41^m41^s.6$, DEC(2000): $-01^{o}55^{'}03^{''}$
is 220 $\mu$G, which compares well with the total field strength 
(ie 87 $\times$ 2 = 174 $\mu$G) obtained from OH Zeeman measurements.
To our knowledge no Zeeman measurements
of the magnetic field strength of the foreground gas exists. 

\section{Discussion and Summary}
\label{sec:sum}

Radio Recombination line transitions H166$\alpha$, C166$\alpha$, X166$\alpha$, H110$\alpha$ and
C110$\alpha$ were imaged towards the starforming region NGC 2024. We deduced the spatial location
and extent of line emitting regions from the images and the LSR velocity structure.
Further, we constructed non-LTE models for RRL emission to obtain the physical properties of the
line forming regions.

\begin{figure}
\includegraphics[height=3.5in, width=4.5in, angle=0]{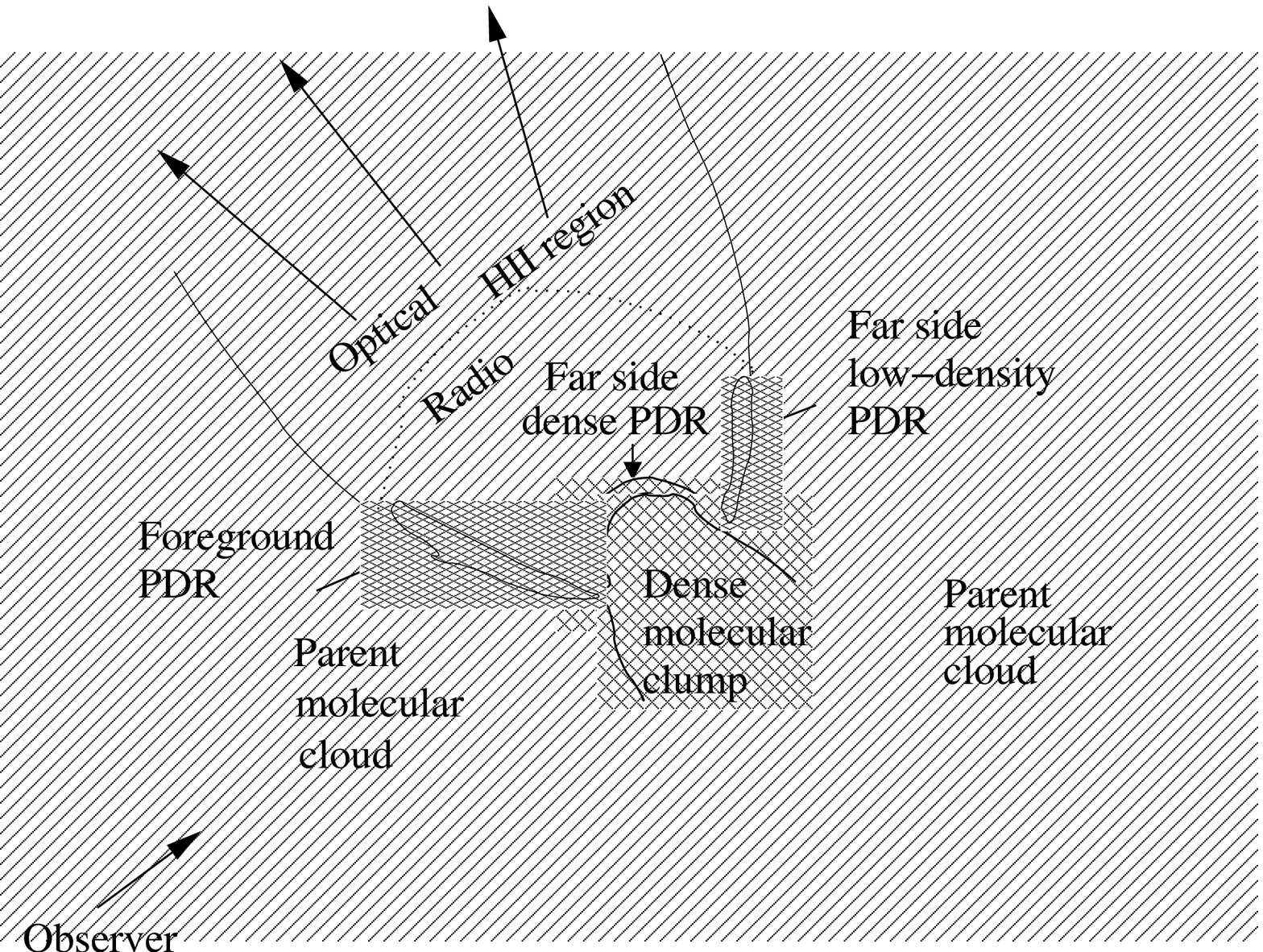}
\caption{
Schematic of the \HII /molecular cloud toward NGC 2024.
This schematic is based on the structure of NGC 2024 proposed by
Barnes \etal (1989) and Kr\"{u}gel \etal\ (1982).
The starforming region comprises of \HII\ region partially enclosed by the
parent molecular cloud with mean density $\sim$ 10$^5$ \cmthree. A dense molecular clump
is located at the far side of the \HII\ region. Photo dissociation regions
exist at the \HII--molecular gas interface. To explain the observed CRRL emission
and OH absorption feature at 10 \kms\ (Crutcher \etal 1999), we suggest that the 
far side dense PDR protrudes into the \HII\ region. 
\label{fig5} } 
\end{figure}

A schematic of NGC 2024 is shown in Fig.~\ref{fig5}.
This schematic is based on the structure of NGC 2024 proposed by
\nocite{betal89}Barnes \etal (1989) and \nocite{ketal82}Kr\"{u}gel \etal\ (1982).
As illustrated, the starforming region comprises of \HII\ region partially enclosed by the 
parent molecular cloud with mean density $\sim$ 10$^5$ \cmthree. A dense molecular clump, 
with core density larger than $\sim$ 10$^6$
\cmthree, is located at the far side of the \HII\ region. Photo dissociation regions 
exist at the \HII--molecular gas interface. We refer to the interface 
between the parent molecular cloud and \HII\ region at the near
side as the `foreground PDR' and that at the far side as the `far side low-density PDR'. 
The interface between the dense clump and \HII\ region
is referred to as the `far side dense PDR'.  

The C166$\alpha$ line image reveals that the emission extends in the N-S direction. The LSR velocity of the line ranges from $\sim$ 8 \kms\ in the N to $\sim$ 10 \kms\ in the SE,
with a mean velocity of 9 \kms. The mean velocity is similar to that of
molecular lines observed from the near side of the \HII\ region. Thus we
conclude that the C166$\alpha$ emission originates in the foreground PDR.
Non-LTE modeling of CRRL emission shows that this PDR has properties
$T_{PDR} \sim 100$ K, $n_e \sim 5$ \cmthree, $n_H \sim 1.7 \times 10^4$ \cmthree and
$l \sim 0.06$ pc. The model also shows that in the absence of stimulated emission,
CRRL emission from the far side low-density PDR is not detectable. 
The estimated magnetic field in the PDR is $\sim$ 60 $\mu$G. 

The image of C110$\alpha$ line shows that the emission is confined close to the
southern boundary of the \HII\ region with an LSR velocity of 10.3 \kms. This
LSR velocity is similar to that of  molecular lines observed from the dense clump at the far side.
We conclude that the C110$\alpha$ originates from the far side dense PDR.
From CRRL modeling we obtain properties for this PDR i.e
$T_{PDR} \sim$ 200 K, $n_e \sim$ 50 \cmthree, $n_H \sim 1.7 \times 10^5$
\cmthree and $l \sim$ 0.03 pc. 
To explain the observed CRRL emission we postulate that 
the far side dense PDR protrudes
into the \HII\ region (see Fig.~\ref{fig5}). We suggest that this protrusion 
is due to Lyman continuum photons
eroding the low density molecular material on either side of the dense clump. 
This geometry is consistent with the detection of OH absorption line 
at 10 \kms, since part of the radio continuum originates from behind
the far side dense PDR (\nocite{cetal99}Crutcher \etal 1999). The estimated 
magnetic field in this PDR is $\sim$ 220 $\mu$G, comparable
with field strength obtained from OH Zeeman observations (\nocite{cetal99}Crutcher \etal 1999).

The H166$\alpha$ line exhibits narrow and broad components. The LSR velocity
of narrow hydrogen is $\sim$ 9 \kms, similar to that of molecular lines in the foreground gas
and C166$\alpha$. Images of H$^0$166$\alpha$ shows that the emission extends in the N-S direction
and is similar to C166$\alpha$ emission. We conclude that the H$^0$166$\alpha$
line emission is co-located with the foreground PDR. The gas properties
deduced from modeling H$^0$166$\alpha$ line emission are 
$T_{PDR} \sim 50 $ K, $n_e \sim 4$ \cmthree and $l \sim 0.01$ pc. 
The ionization fraction in the H$^0$ line forming region is $\sim$ 10$^{-4}$.
The broad H166$\alpha$ line originates from the \HII\ region.

A recombination line from a heaver
element has been observed at 1.4 GHz. The spatial distribution
of the X166$\alpha$ is different from that of H$^0$166$\alpha$ and
C166$\alpha$ line emission. The smaller line width of X166$\alpha$
compared to C166$\alpha$ line may indicate that the line
emission is associated with colder dust clouds. Based on the
expected low depletion of sulphur in such clouds
and the $-$8.1 \kms\ velocity offset relative to C166$\alpha$,
we conclude that the X166$\alpha$ line emission arises from sulphur.

Observations presented in this paper are the first high-angular
resolution RRL observations with full spatial information at 4.8 GHz.
The 1.4 GHz observations complement earlier interferometric observations
presented by \nocite{aetal90}Anantharamaiah \etal (1990). Our data
set was used to constrain non-LTE models of CRRL and H$^0$RRL line
emission. The high angular resolution of the observations allowed
us to measure the angular size and construct models for the line forming region.
Based on this modeling we have presented
a new geometry for the \HII /molecular cloud toward NGC 2024 (see
Fig~\ref{fig5}). We have also used the data
set to demonstrate for the first time that the magnetic field
obtained using CRRLs is comparable with values obtained from
Zeeman observations.

\acknowledgments
We are grateful to the anonymous referee for the critical
comments and suggestions which have helped in
refining the interpretation of our observations and also
significantly improved the paper.
We thank F.J. Lockman for his help and advice during data reduction,
Glen Langston for sharing his computer code to convert GBT data format to
UVFITS fits and Dana Balser for his useful comments on the manuscript.


\begin{deluxetable}{lrr}
\tabletypesize{\small}
\tablecolumns{3}
\tablewidth{0pc}
\tablecaption{Observing log \label{tab1}}
\tablehead{
\colhead{Parameters} & \multicolumn{2}{c}{Values} \\ 
\colhead{}           & \colhead{166$\alpha$}  & \colhead{110$\alpha$} }
\startdata
\cutinhead{VLA observations} 
Date of observations       & 08-SEP-2000  &  01-AUG-2000  \\
Field center RA (J2000)    & $05^h41^m44.5^s$ &  $05^h41^m44.5^s$     \\
Field center DEC (J2000)   & $-01^{o}54^{'}39^{''}$ & $-01^{o}54^{'}39^{''}$ \\
RRLs observed         & C166$\alpha$ & C110$\alpha$  \\ 
                     & H166$\alpha$ & H110$\alpha$ \\ 
Velocity range (\kms) -- CRRL & 40        & 47  \\
                      -- HRRL & 162       & 95  \\
Velocity resolution (\kms)-- CRRL  & 1.3    & 1.5   \\
                          -- HRRL & 3.1    & 1.8      \\
Phase calibrator           & J0521+166    & J0530+135 \\
On-source observing time (hrs) & 2.9      & 2.6   \\
Synthesized beam (arcsec)      & 75 $\times$  67 & 18 $\times$  15   \\
Position angle of  the          &          &            \\ 
synthesized beam (deg)          & $-81^o$  & $-41^o$  \\
Largest angular size (arcmin)   & 20      & 6.5  \\
RMS noise in the spectral       &   &    \\
cube (mJy/beam)                 & 8 & 15 \\
RMS noise in the                &   &  \\ 
continuum images (mJy/beam)     & 4  & 22   \\
\cutinhead{GBT observations} 
Date of observations          &\nodata & 12-AUG-2002  \\
Field center RA (J2000)       &\nodata & 05$^h$39$^m$13$^s$.0     \\
Field center DEC (J2000)      &\nodata & $-$01$^{o}$56$^{'}$04$^{''}$   \\
RRLs observed                 &\nodata  & C110$\alpha$, H110$\alpha$   \\ 
Velocity range (\kms)         &\nodata & 308  \\
Velocity resolution (\kms)    &\nodata & 0.3   \\ 
On-source observing time (hrs)&\nodata & 1.0   \\
Beam (arcmin)                 &\nodata & 2.7 $\times$ 2.7   \\
RMS noise in the spectral     &\nodata &    \\
cube (mJy/beam)               &\nodata & 34  \\
RMS noise in the              &\nodata &  \\ 
continuum images (mJy/beam)   &\nodata &  12 \\
\enddata
\end{deluxetable}

\begin{deluxetable}{lllrrrr}
\tabletypesize{\small}
\tablecolumns{7}
\tablewidth{0pc}
\tablecaption{Observed parameters \label{tab2}}
\tablehead{
\colhead{Line} & \multicolumn{2}{c}{Flux density} & \colhead{V$_{LSR}$} & \colhead{$\Delta V$\tablenotemark{a}} & \colhead{Size\tablenotemark{b}} & \colhead{Comment} \\
               & \colhead{cont} &\colhead{line} &  & & & \\
               & \colhead{(Jy)} & \colhead{(Jy)}   & \colhead{(\kms)} & \colhead{(\kms)} & \colhead{(\arcsec $\times$ \arcsec)} &  
}
\startdata
\cutinhead{HII region}
H166$\alpha$ & 7.1 & 0.044(0.004) &  3.3(1.2)   &  33.4(2.8) & 76 $\times$ 76 & Note c\\
H110$\alpha$ & 6.9 & 0.444(0.006) &  5.7(0.1)   &  22.6(0.3) & 76 $\times$ 76 & Note c\\
\cutinhead{H$^0$ line}
H$^0$166$\alpha$ & 7.1 & 0.09(0.01\tablenotemark{g})     &  9.0        &  1.7(1.5) &76 $\times$ 76 & Note c\\
H$^0$110$\alpha$ & 6.9 & (0.02)       & \nodata     & \nodata   &76 $\times$ 76 & Note c\\
\cutinhead{Far side carbon line emission}
C166$\alpha$ & 7.1 & 0.08(0.01)\tablenotemark{f} &  8.8(0.1) &  2.4(0.2) &76 $\times$ 76 & Note c\\
C110$\alpha$ & 6.9 & 0.10(0.01) & 10.3(0.2) &  2.8(0.4) &76 $\times$ 76 & Note c\\ 
\cutinhead{Foreground carbon line emission}
C166$\alpha$ & 22.3 & 0.18(0.02) &  8.8(0.2) &  2.5(0.3) & 236 $\times$ 108 & Note d\\
C110$\alpha$ & 20.1 & (0.06)     & \nodata   & \nodata   & 236 $\times$ 108 & Note d\\
\cutinhead{X region}
X166$\alpha$ & 22.3 & 0.10(0.03) &  0.7\tablenotemark{e}(0.2) &  0.8(0.4) & 236 $\times$ 108 & Note d\\
X110$\alpha$ & 20.1 & (0.06)    &  \nodata      & \nodata       & 236 $\times$ 108 & Note d\\
\enddata

\tablenotetext{a}{Line widths are corrected for the broadening due to finite spectral resolution.}
\tablenotetext{b}{Size of the region over which the data are averaged to obtain the spectra.}
\tablenotetext{c}{Spectra averaged over Region B shown in Fig. 2. The region is centered at RA(2000): $05^h41^m41^s.6$, DEC(2000): $-01^{o}55^{'}03^{''}$.}
\tablenotetext{d}{Spectra averaged over Region A in Fig. 2, The region is centered at RA(2000): $05^h41^m42^s.7$, DEC(2000): $-01^{o}53^{'}55^{''}$.}
\tablenotetext{e}{LSR velocity with respect to carbon.}
\tablenotetext{f}{The C166$\alpha$ line flux density can have a contribution from foreground PDR and therefore the listed flux density should be considered as an upper limit to the emission from the far side PDR.}
\tablenotetext{g}{The error in amplitude is estimated from the residual spectrum obtained after subtracting the Gaussian components}

\end{deluxetable}

\begin{thebibliography}{}
\bibitem[Anantharamaiah et al. 1990]{aetal90}
Anantharamaiah, K. R., Goss, W. M. \& Dewdney, P. E. 1990, in Proceedings of IAU Colloq. 125,
Radio Recombination Lines: 25 Years of Investigation, ed. M. A. Gordon \& R. L. Sorochenko (
Puschino, USSR: Dordrecht:Kluwer), 123

\bibitem[Anthony-Twarog 1982]{a82}
Anthony-Twarog, B. J. 1982, AJ, 87, 1213

\bibitem[Ball et al. 1970]{betal70}
Ball, J. A., Cesarsky, D., Dupree, A. K., Goldberg, L., \& Lilley, A. E. 1970,
ApJ, 162L, 25

\bibitem[Barnes et al. 1989]{betal89}
Barnes, P. J., Crutcher, R. M., Bieging, J. H., Storey, J. W. V., \& Willner, S. P. 1989, ApJ, 342, 883

\bibitem[Bik et al. 2003]{betal03}
Bik, A., Lenorzer, A., Kaper, L., \etal 2003, A\&A, 404, 249 

\bibitem[Brocklehurst \& Salem 1977]{bs77}
Brocklehurst, M. \& Salem, M. 1977, Computer Phys. Commun., 13, 39

\bibitem[Buckle et al. 2010]{betal10}
Buckle, J. V., Curtis, E. I., Roberts, J. F., \etal 2010, MNRAS, 401, 204

\bibitem[Chaisson 1973]{c73}
Chaisson, E. J. 1973, ApJ, 182, 767

\bibitem[Crutcher et al. 1999]{cetal99}
Crutcher, R. M., Roberts, D. A., Troland, T. H., \& Goss, W. M. 1999, ApJ, 515, 275

\bibitem[Crutcher 1999]{c99}
Crutcher, R. M. 1999, ApJ, 520, 706

\bibitem[Crutcher et al. 1986]{cetal86}
Crutcher, R. M., Henkel, C., Wilson, T. L., Johnston, K. J., \& Bieging, J. H. 1986, ApJ, 307, 302

\bibitem[Dupree 1974]{d74}
Dupree, A. K. 1974, ApJ, 187, 25

\bibitem[Gordon 1969]{g69}
Gordon, M. A. 1969, ApJ, 158, 479

\bibitem[Graf et al. 1993]{getal93}
Graf, U. U., Eckart, A., Genzel, R., \etal 1993, ApJ, 405, 249

\bibitem[Graf et al. 2012]{getal12}
Graf, U. U., Simon, R., Stutzki, J., \etal 2012, A\&A, 542L, 16

\bibitem[Henkel et al. 1980]{hww80}
Henkel, C., Walmsley, C. M., \& Wilson, T. L. 1980, A\&A, 82, 41

\bibitem[Hoang-Binh \& Walmsley 1974]{hw74}
Hoang-Binh, D., \& Walmsley, C. M. 1974, A\&A, 35, 49 

\bibitem[Kr\"{u}gel et al. 1982]{ketal82}
Kr\"{u}gel, E., Thum, C., Pankonin, V., \& Martin-Pintado, J. 1982, A\&AS, 48, 345


\bibitem[MacLeod et al. 1975]{metal75}
MacLeod, J. M., Doherty, L. H., \& Higgs, L. A. 1975, A\&A, 42, 195

\bibitem[Mangnum \& Wootten 1993]{mw93}
Mangnum, G. J., \& Wootten, A. 1993, ApJS, 89, 123

\bibitem[Mezger et al. 1992]{metal92}
Mezger, P. G., Sievers, A. W., Haslam, C. G. T., \etal 1992, A\&A, 256, 631

\bibitem[Morton 1974]{m74}
Morton, D. C, 1974, ApJL 193, L35

\bibitem[Natta et al. 1994]{nwt94}
Natta, A., Walmsley, C. M., \& Tielens, A. G. G. M.  1994, ApJ, 428, 209

\bibitem[Palmer et al. 1967]{petal67}
Palmer, P., Zuckerman, B., Penfield, H., Lilley, A. F., \& Mezger, P. G. 1967, AJ, 72, 821  

\bibitem[Pankonin et al. 1977]{petal77}
Pankonin, V., Walmsley, C. M., Wilson, T. L., \& Thomasson, P. 1977, A\&A, 57, 341


\bibitem[Roshi 2007]{r07}
Roshi, D. A. 2007, ApJ, 658L, 41 

\bibitem[Shaver 1975]{s75}
Shaver, P. A. 1975, Parmana, 5, 1

\bibitem[Snell et al. 1984]{setal84}
Snell, R. L, Mundy, L. G., Goldsmith, P. F., Evans II, N. J., \& Erickson, N. R. 1984,
ApJ, 276, 625

\bibitem[Subrahmanyan et al. 1997]{setal97}
Subrahmanyan, R., Goss, W. M., Megeath, S. T., \& Barnes, P. J. 1997, MNRAS, 290, 431

\bibitem[Tucker et al. 1973]{tetal73}
Tucker, K. D., Kutner, M. L., \& Thaddeus, P. 1973, ApJ, 186L, 13

\bibitem[Rickard et al. 1977]{retal77}
Rickard, L. J, Zuckerman, B., Palmer, P., \& Turner, B. E. 1977, ApJ, 218, 659

\bibitem[van der Werf et al. 1993]{vetal93}
van der Werf, P. P., Goss. W. M., Heiles, C., Crutcher, R. M., \& Troland, T. H. 1993, ApJ, 
411, 247

\bibitem[Walmsley \& Watson 1982]{ww82}
Walmsley C. M., \& Watson W. D. 1982, ApJ, 260, 317

\bibitem[Wilson \& Thomasson 1975]{wt75}
Wilson, T. L., \& Thomasson, P. 1975, A\&A, 43, 167 

\end{thebibliography}
\end{document}